\documentclass[review]{elsarticle}

\usepackage{hyperref}
\usepackage{graphicx}
\usepackage{bmpsize}

\journal{Journal of \LaTeX\ Templates}

\begin{document}

\begin{frontmatter}

\title{Measurements of D-D fusion neutrons generated in nanowire array laser plasma using Timepix3 detector}

\author[UTEFaddress]{Peter Rubovi\v c\corref{mycorrespondingauthor}}
\cortext[mycorrespondingauthor]{Corresponding authors}
\ead{peter.rubovic@cvut.cz}



\author[AMaddress,INFNaddress]{Aldo Bonasera}	
\author[UTEFaddress,PILSENaddress]{Petr Burian}	
\author[BEIJINGaddress]{Zhengxuan Cao}	
\author[MOEaddress]{Changbo Fu}	
\author[BEIJINGaddress]{Defeng Kong}	
\author[HENGYANGaddress]{Haoyang Lan}	
\author[SIAPaddress]{Yao Lou}	
\author[HENGYANGaddress]{Wen Luo}	
\author[CIAEaddress]{Chong Lv}	
\author[SIAPaddress,MOEaddress]{Yugang Ma}	
\author[BEIJINGaddress]{Wenjun Ma\corref{mycorrespondingauthor}}	
\ead{wenjun.ma@pku.edu.cn}
\author[HENGYANGaddress]{Zhiguo Ma}	
	\author[UTEFaddress]{Luk\'a\v s Meduna}	
\author[BEIJINGaddress]{Zhusong Mei}	
\author[UTEFaddress]{Yesid Mora}	
\author[BEIJINGaddress]{Zhuo Pan}	
\author[BEIJINGaddress]{Yinren Shou}	
\author[UTEFaddress]{Rudolf S\'ykora}	
\author[UTEFaddress]{Martin Veselsk\'y}	
\author[BEIJINGaddress]{Pengjie Wang}	
\author[INPACaddress]{Wenzhao Wang}	
\author[BEIJINGaddress]{Xueqing Yan}	
\author[SIAPaddress,SARIaddress]{Guoqiang Zhang}	
\author[BEIJINGaddress]{Jiarui Zhao}	
\author[BEIJINGaddress]{Yanying Zhao}		
\author[UTEFaddress]{Jan \v Zemli\v cka}	

\address[UTEFaddress]{Czech Technical University in Prague, Institute of Experimental and Applied Physics, Husova 240/5, 110 00 Prague 1, Czech Republic}
\address[AMaddress]{Cyclotron Institute, Texas A\&M University, College Station, Texas 77843, USA}
\address[INFNaddress]{Laboratori Nazionali del Sud, INFN, via Santa Sofia, 62, 95123 Catania, Italy}
\address[PILSENaddress]{University of West Bohemia, Faculty of Electrical Engineering, Univerzitni 8, Pilsen, Czech Republic}
\address[BEIJINGaddress]{State Key Laboratory of Nuclear Physics and Technology, School of Physics, Peking University, Beijing 100871, China}
\address[MOEaddress]{Key Laboratory of Nuclear Physics and Ion-Beam Application (MOE), Institute of Modern Physics, Fudan University, Shanghai 200433, China}
\address[HENGYANGaddress]{School of Nuclear Science and Technology, University of South China, Hengyang 421001, China}
\address[SIAPaddress]{Shanghai Institute of Applied Physics, Chinese Academy of Sciences, Shanghai 201800, China}
\address[CIAEaddress]{China Institute of Atomic Energy, P. O. Box 275(10), Beijing 102413, China}
\address[INPACaddress]{INPAC and School of Physics and Astronomy, Shanghai Jiao Tong University, Shanghai 200240, China}
\address[SARIaddress]{Shanghai Advanced Research Institute, Chinese Academy of Sciences, Shanghai 201210, China}

\begin{abstract}
We present the results of neutron detection in a laser plasma experiment with a CD$_2$ nanowire target. A hybrid semiconductor pixel detector Timepix3 covered with neutron converters was used for the detection of neutrons. D-D fusion neutrons were detected in a polyethylene converter through recoiled protons. Both the energy of recoiled protons and the time-of-flight of neutrons (and thus their energy) were determined. We report $(2.4 \pm 1.8) \times 10^7$ neutrons generated for 1~J of incoming laser energy. Furthermore, we proved that Timepix3 is suitable for difficult operational conditions in laser experiments.
\end{abstract}


\end{frontmatter}


\section{Introduction}

Ever since the invention of laser, a source of coherent light, continuous effort has been spent on increasing the peak power and reaching still higher energy density. The tremendous improvements in laser technologies at relatively low costs open up many possibilities not only in practical applications in medicine, energy production, etc., but also in fundamental physics applications \cite{Strickland1985, Remington1999, Atzeni2004, Nuckolls1972, Lindl1995, Bonasera2008}. Using lasers, it is possible to create plasmas of different kinetic energy densities, which can be measured very precisely. In such systems, nuclear fusions can occur, which in turn might yield important information about the dynamics of plasmas. It is not important at all that the plasma is in equilibrium \cite{Bonasera2008}: such condition is useful to perform analytical estimates of the processes occurring in the plasma. In nature, many of the processes do not necessarily go through equilibrium phases and this is also true in laser-matter interactions where an unstable system is formed for a very short time \cite{Labaune2013}. Nuclear fusion reactions in these plasmas could be enhanced if the kinetic energy distribution of the system was tailored to a region where resonances are present in the compound nucleus. A delicate balance between the available laser energy and the kinetic energy of the produced ions might be reached by a suitable choice of the target properties (density, thickness, composition, etc.), laser focalization on the target, impulse duration, etc. It is important that most of the laser energy is transferred to the highest possible number of ions with an optimized kinetic energy distribution. If the efficiency of energy transfer from the lasers to the ions is low, high electromagnetic pulse and X-rays are generated, which are a great disturbance to the experimental devices. In such highly non-equilibrium scenarios one can not only still derive physical quantities such as fusion cross sections relevant for instance in nuclear astrophysics \cite{Bonasera2008, Bang2013, Barbui2013, Lattuada2016}, but also derive efficient methods for energy production, radioisotopes production \cite{Ledingham2004} and other features very useful in medicine and other fields as well.

The availability of high-intensity laser facilities delivering petawatts of power into a very small volume opened the possibility to perform investigations related to fundamental and applied nuclear physics. In particular, nuclear astrophysics is bound to benefit from this opportunity. The high-intensity laser beams allow to accelerate nuclear particles to energies, enabling us to observe nuclear reactions in the plasma, simulating the conditions in astrophysical objects such as stars at various stages of their lifecycle. Several mechanisms of particle acceleration can be used. The target normal sheath acceleration (TNSA) is the mechanism where the intense laser beam impinges onto solid target and the gas of hot electrons after passing through the thin foil accelerates positively charged ions, predominantly protons. Such protons can reach energy of tens of Mega-electron-Volts (MeV) and can be used as excellent probe for evolution of plasma. TNSA protons can be also used for investigation of proton-induced nuclear reactions. The reaction of protons with boron isotopes, which can be used in nuclear fusion reactors for heating of plasma is of particular interest and it is important also for the study of nucleosynthesis in astrophysical objects. The radiation pressure acceleration (RPA) is the mechanism where intense laser beam hits ultra-thin foil and the beam of neutral particles with solid state density can be produced.  It is expected to produce protons with GeV energies and also beams of heavy ions with energies sufficient for initiation of nuclear reactions. 

The production of energy using nuclear fusion initiated by lasers is one of the primary goals of several high-power laser facilities such as NIF, Omega, Shen-Guang II and III, etc. \cite{Zhang2019}. Usually, a multiplicity of synchronous laser beams are fired into a central position where a pellet, containing deuterium and tritium, is located and fusion is initiated by inertial collapse of the pellet and formation of hot plasma. A successful ignition has not been observed so far. Success of such endeavor depends on many factors, such as specific construction of pellets and fusion cross sections in the hot plasma. It is expected due to electron screening \cite{Assenbaum1987} that cross sections of low energy nuclear reactions in a dense plasma will be higher than the values measured in standard nuclear experiments, where a beam of particles hits the solid state or gaseous target. The screening from the surrounding electronic plasma can lead to an increase of Coulomb barrier penetration probabilities and thus to an increase of overall cross sections. Also, the plasma conditions can influence the energy losses of the moving charged particles and therefore increase the probability of nuclear reactions. Experiments at the high-power lasers offer unique opportunity to investigate such unusual effects and describe them quantitatively, thus benefiting both basic science, such as the study of nucleosynthesis, and applied science, such as nuclear fusion research and development. 

Recent work of Curtis et al. \cite{Curtis2018} demonstrated that the consistency of the target irradiated by laser beam can dramatically influence the efficiency of conversion of laser energy into energy of hot dense plasma. Use of nano-structured deuterium loaded target appears to rapidly increase the production of accelerated deuterium ions, leading to record number of fusion reactions per joule of laser energy. This is an important step towards production of energy using laser initiated fusion. The experiment, staged at the Compact Laser Plasma Accelerator (CLAPA) facility at the Peking University in Beijing, aimed at verifying this result. In comparison with experiment of Curtis et al., the Timepix3 hybrid pixel semiconductor detector was used for detection of neutrons and for the determination of their time-of-flight (TOF). In comparison with routinely used scintillation detectors, neutrons can be identified using the characteristic tracks of recoiled protons, scattered in the polyethylene, covering the silicon sensor. It was aimed to test the Timepix3 detectors under conditions of high electro-magnetic noise. 

The CLAPA facility is equipped with a 200 terawatt laser with typical energy in the pulse of several joules and pulse duration of few tens of femtoseconds. The Timepix3 detector was placed just outside of the target chamber, at a distance of $\sim 1$~m from the target position. The CD$_2$ nanowire array targets were prepared by extrusion of deuterated polyethylene into the Anodized Aluminum Oxide template (AAO). The average density of such target is $0.1-0.3\,\textrm{g}\cdot \textrm{cm}^{-3}$, determined by measuring the void ratio with scanning electron microscope, which is lower than the usual solid density ($0.9\,\textrm{g}\cdot \textrm{cm}^{-3}$). Under the assumption that the target is completely ionized, the electron density would be about $10^{22}\,\textrm{cm}^{-3}$,which is significantly higher than the critical electron density ($n_\textrm{c} = \frac{m_\textrm{e} \omega_\textrm{L}^2}{4 \pi e^2} = 1.7 \times 10^{21} \, \textrm{cm}^{-3}$, where $m_\textrm{e}$, $\omega_\textrm{L}$, and $e$ are the mass of electron, the frequency of the laser, and the charge of electron, respectively). Laser can propagate into the array with a high absorption efficiency up to nearly 100\% \cite{Curtis2018}.

This paper describes the detection of D-D fusion neutrons with 2.45 MeV energy by the Timepix3 hybrid semiconductor pixel detector. The paper is divided into 4 parts. After Introduction, the experimental configuration, the detection techniques and the data analysis are described in Materials and Methods. Afterwards, the obtained results are shown and discussed.

\section{Materials and methods}
\paragraph{Laser system}

\begin{figure}[h]
			\centering
			 \includegraphics[width=7cm]{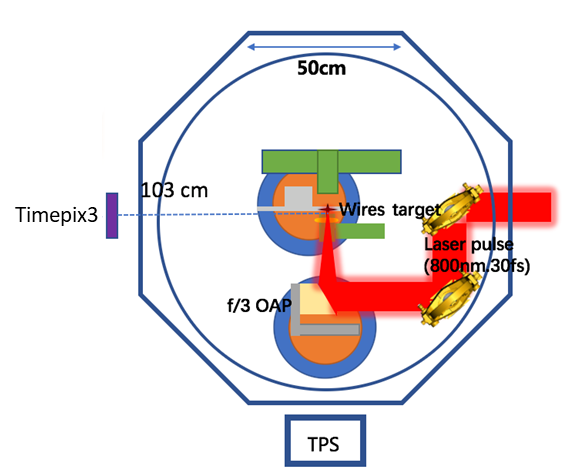}
			 \caption{The layout of the experimental chamber as described in the text. TPS stands for Thomson parabola spectrometer.} 
			\label{figure1}
\end{figure}

The experiment was conducted by using the 30~fs, 200~TW CLAPA laser system at Peking University. To prevent the pre-expansion \cite{Zeng2005, Doggett2011} of the nanowires caused by the pre-pulses \cite{Wang2018, Doumy2004, Ovchinnikov2013, Batani2010}, a plasma mirror (PM) system was employed to improve the temporal contrast \cite{Doumy2004} of laser by two orders of magnitude, ensuring that the pre-pulse intensity is below $10^{11}$~W$\cdot$cm$^{-2}$ at 5~ps \cite{Geng2018}. The on-target energy is 0.8~J. The laser beam was focused into a spot with FWHM of 4.5~$\mu$m diameter in order to reach intensities up to $3.8 \times 10^{19}$~W$\cdot$cm$^{-2}$ by using an f/3, 90 degree off-axis-parabolic (OAP) mirror. Refer to Figure~\ref{figure1} for the illustration of the experimental chamber.

\paragraph{CD$_2$ nanowire arrays targets}

Ordered high-aspect-ratio deuterated polyethylene (DPE) nanowires array targets were used in the experiment. They were prepared by the following process: A thick DPE sheet was attached to the Anodized Aluminum Oxide template (AAO) \cite{Wang2003}, which has dense hexagonal nanopores and is obtained by anodizing aluminum in the electrolyte. By heating and mechanically compressing the sheet, the DPE molecules are driven into the templates. Afterwards, the DPE nanowire arrays were obtained by dissolving the AAO membrane in 2M NaOH solution for 30 min \cite{Zhang2001}.
 
\begin{figure}[h]
			\centering
			 \includegraphics[width=7cm]{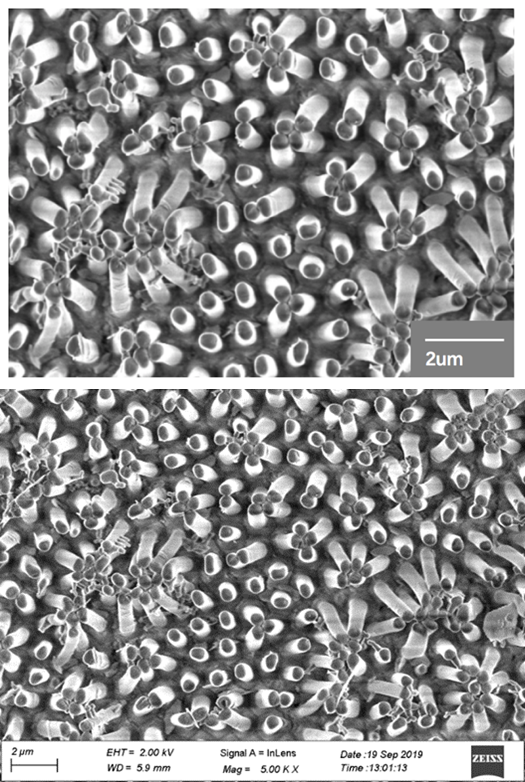}
			 \caption{The SEM image of the DPE nanowires array target with 500~nm diameter, 800~nm spacing and 5~$\mu$m length.} 
			\label{figure2}
\end{figure}

The diameters, spacings and lengths of the nanowires were determined by the templates. By varying the temperature of melting and the time of dissolving, DPE nanowire arrays with different morphology were prepared. A SEM image of a typical DPE nanowires array target is shown in Figure~\ref{figure2}. The thickness of the substrate that supports the nanowire arrays is about 500~$\mu$m.

\paragraph{Timepix3 detector}

We used the hybrid semiconductor pixel detector Timepix3 \cite{Poikela2014} for the current experiment. The chip is divided into the $256 \times 256$-pixel matrix (65,536 pixels in total covering the area of 1.96~cm$^2$) with pixel pitch equal to 55~$\mu$m. The detector can be read in a frame driven mode like earlier Medipix/Timepix family detectors or in a sparse data driven mode where each pixel is read out immediately after exceeding the threshold level with a per-pixel deadtime of 475~ns for hit rates up to 40~MHits$\cdot$cm$^{-2}$$\cdot$s$^{-1}$. The following information is available simultaneously for each pixel hit: its coordinates, deposited energy, and time of the hit with time resolution of 1.56~ns. The average energy threshold for Timepix3 detectors is as low as 1.8~keV, however we used a conservative value of 3.6~keV for the experiment. The Gigabit Ethernet embedded readout interface Katherine was used for the detector operation \cite{Burian2017}. The range of possible applications is reviewed in \cite{Ballabriga2018}. Such detector has already been used for measuring neutron TOF \cite{Bergmann2016} at higher energies, where nuclear reactions in the sensor material happen with considerably longer TOF base.

\paragraph{Experimental arrangement}

The Timepix3 detector with a 500~$\mu$m thick silicon (Si) sensor was used in the actual experiment. In order to detect neutrons, the surface of the sensor was covered with several neutron converters. Approximately one quarter of the sensor was covered with 90\% $^6$Li enriched LiF sprayed onto 50~$\mu$m thick aluminum for the detection of thermal neutrons, where LiF layer was facing the detector. In this way, products of $^6$Li(n, $\alpha$)$\tau$ reaction can be detected. One third of the sensor was covered with 1 mm thick polyethylene (PE) converter for the detection of fast neutrons through recoiled protons (hydrogen nuclei) created in elastic scattering of incident neutrons. For data analysis, the area under PE converter covering $94\times256$ pixels (corresponding to $\sim$73~mm$^2$) was taken. The rest of the sensor was left uncovered in order to distinguish (i) background signal, (ii) non-neutron induced signal, and (iii) signal induced by fast neutrons with energy higher than $\sim$4~MeV through nuclear reaction directly in the Si sensor \cite{Rubovic2018}.

The detector was placed outside the interaction chamber with converters facing the chamber. The initial distance from the laser target was 148~cm, later the detector was moved closer to a distance set to 103~cm. For details on the experimental layout, see Figure~\ref{figure1}. During the experiment, 81 shots were fired in total.

\paragraph{Detection principles \& data analysis}

\begin{figure}[h]
			\centering
			 \includegraphics[width=12cm]{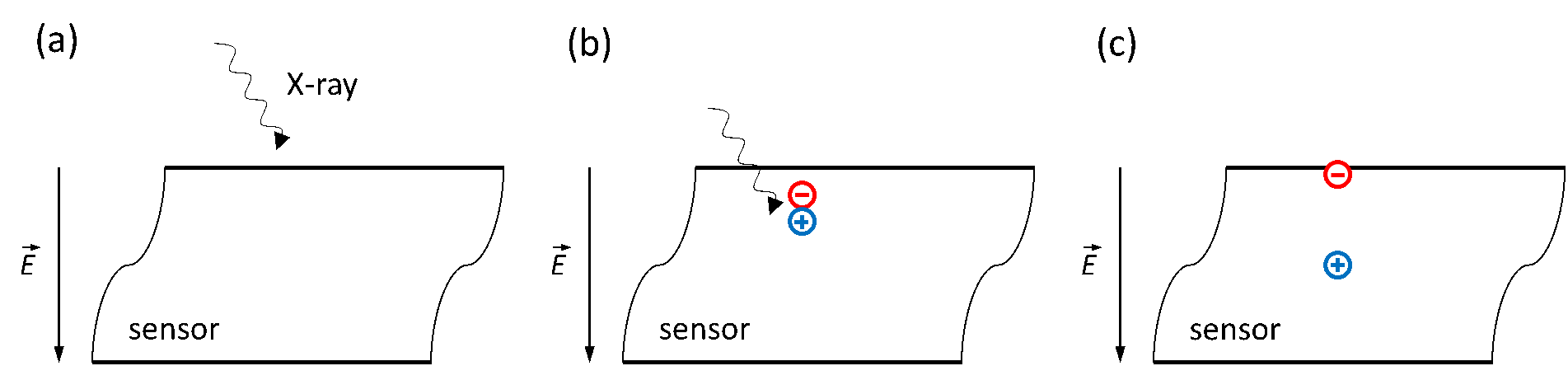}
			 \caption{Schematics of the X-ray detection in a semiconductor detector. The incident particle is absorbed by photoelectric effect in a few microns’ depth. Subsequently, electron-hole pairs are created, and they drift in opposite directions according to the direction of applied electric field. Low energy (up to a few tens of keV) electrons are detected in a similar way.} 
			\label{figure3}
\end{figure}

\begin{figure}[h]
			\centering
			 \includegraphics[width=12cm]{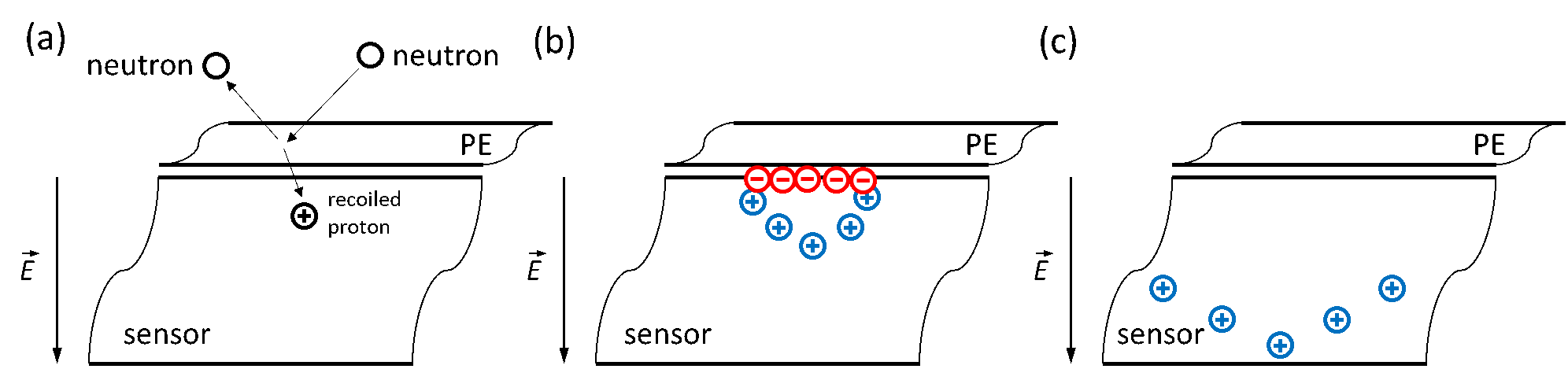}
			 \caption{Schematics of the fast neutron detection in a semiconductor detector using hydrogen-rich materials, in our case polyethylene. First, neutrons undergoe elastic collision on hydrogen. Then the recoiled proton leaves the PE converter and enters the Si sensor. The proton is then absorbed in a few microns’ depth (the depth for a proton with 1 MeV energy is around 17~$\mu$m). A large number of electron-hole pairs is created and they expanded due to the charge repulsion. Holes are detected during broad, several tens ns time intervals.} 
			\label{figure4}
\end{figure}

In order to determine neutron energy through TOF measurement, the following steps have to be performed: (i) recoiled protons have to be identified; (ii) the time stamp corresponding to its arrival has to be assigned; (iii) starting point for the TOF measurement, preferably originating from low energy X-ray or electron has to be identified and (iv) its time stamp assigned. Mechanisms of low energy X-ray/electron and fast neutron detection with semiconductor detectors are explained in Figures~\ref{figure3} and \ref{figure4}, respectively. The mean free path for a 4~keV photon in Si is $\sim$10~$\mu$m \cite{Seltzer1987}. Similarly, the range of a 50 keV electron in Si is $\sim$24~$\mu$m \cite{Seltzer1993}. Upon their absorption, electron-hole pairs are created and charge carrying particles are drifting to the opposite sides of the detector. Since we used p-in-n planar Si sensor configuration, holes are attracted to the frontside and their charge is processed by the pixelated ASIC. Since the overall charge is several tens of keV at maximum, it is localized mostly within one or two pixels. The situation is a bit different for fast neutrons detected via recoiled protons. The incident neutron undergoes elastic collision in the hydrogen rich converter layer (in our case PE) and the recoiled proton is detected. The range of protons with energy of 1~MeV is $\sim$17~$\mu$m \cite{Seltzer1993}. Upon its absorption, electron-hole pairs are created. But the large charge density causes their repulsion, thus a large blob covering area of several tens of pixels over several tens ns time interval is detected. The following facts are important for this specific TOF application: For the energy values relevant for the experiment, the depth of absorption is, first, roughly the same for X-rays, electrons, and protons as well. Secondly, the depth is very close to the sensor surface, thus we can assume that the detected holes drift through the entire sensor volume.

\begin{figure}[h]
			\centering
			 \includegraphics[width=7cm]{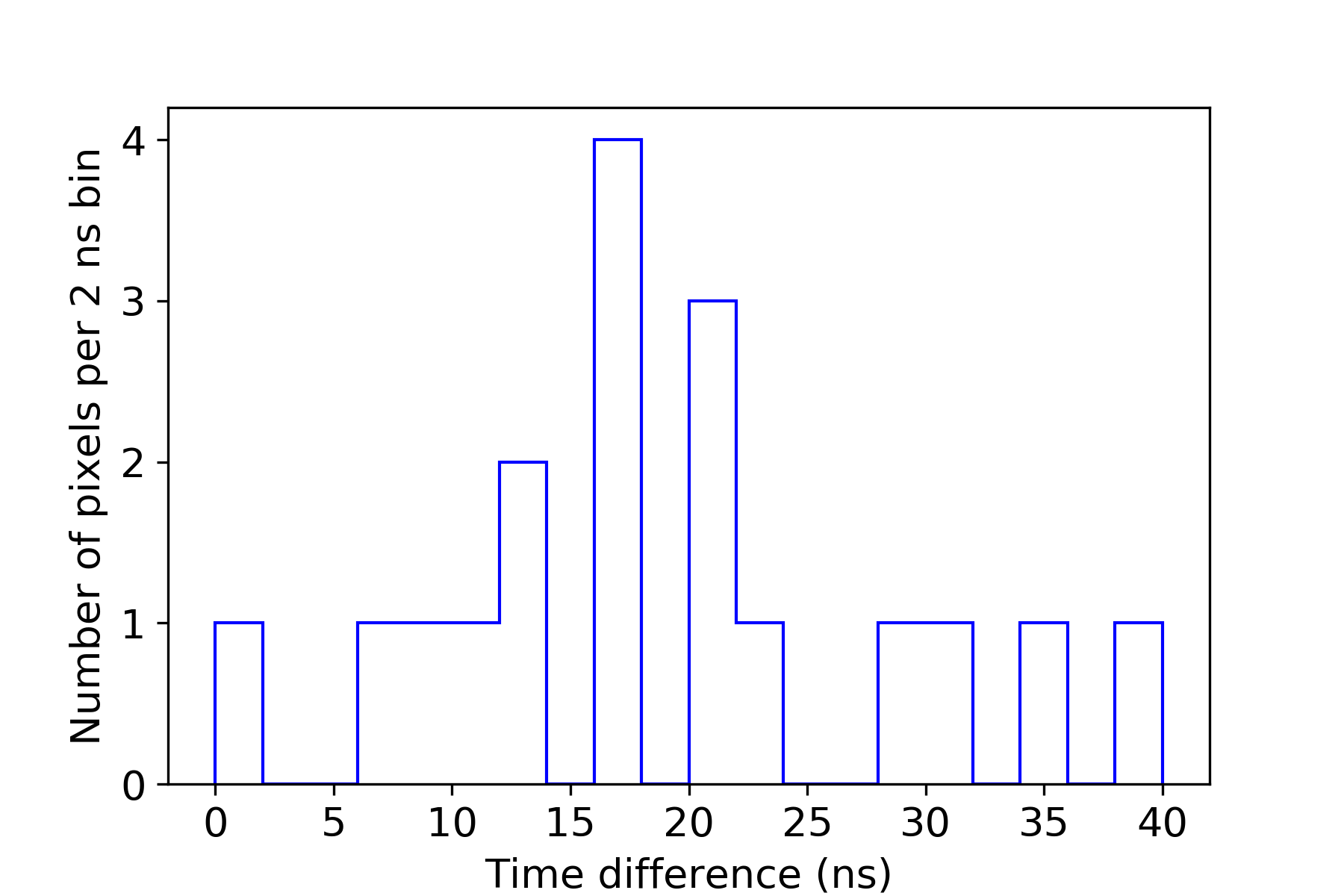}
			 \caption{The time structure of a proton induced event registered by Timepix3 detector.} 
			\label{figure5}
\end{figure}

Following the aforementioned detection principles, the time stamp for low energy X-ray/electron events can be deducted as follows. The time stamp assigned by the detector is related to the detection at the frontside of the sensor, i.e. at the ASIC side. However, the time stamps of X-ray/electron and proton, respectively, must be assigned in the same depth of the sensor. For X-ray/electron events which cover an area of a few pixels, we probe the time stamp from the backside of the sensor (the outer one). For a correct assignment of the proton time stamp we must take into account the time and space structure of a proton track in the Timepix3 detector. A sample time structure of the proton induced event can be seen in Figure~\ref{figure5}.

In order to probe proton induced events from the same depth as in the case of X-ray/low energy electron, we need to take the time stamp from the end of time distribution, see Figure~\ref{figure5}. It is only in this case, when we can subtract the same drift time and to obtain the time stamp corresponding to roughly the same depth in the sensor. In addition, we used the time-walk correction technique for the time stamp accuracy improvement \cite{Bergmann2017}.

\paragraph{Time-of-flight calculation}

To obtain the correct TOF of the detected neutrons we need the following information: distance of the detector from the CD$_2$ target $l$; drift time $t_{\textrm{d}}$; the time stamp of a recoiled proton $t_\textrm{p}$, and the time stamp of an X-ray/electron, which was created during the laser shot $t_\textrm{l}$. It is important to note that the time scale of the laser shots is several orders of magnitude shorter than the time resolution of our detection system, so we can assume that a laser shot happens instantly. Speed of light is denoted as $c$, speed of neutron as $v_\textrm{n}$, and the beginning of the measurement as $t_0$. The time stamp of a recoiled proton can be written as:

\begin{equation}
	t_\textrm{p} = t_0 + \frac{l}{v_\textrm{n}} + t_\textrm{d} \,.
	\label{equation1}
\end{equation}
Similarly, the time stamp of the X-ray can be written as follows:

\begin{equation}
	t_\textrm{l} = t_0 + \frac{l}{c} + t_\textrm{d} \,.
	\label{equation2}
\end{equation}
The speed of the neutron and thus its energy can be derived from equations~\ref{equation1} and~\ref{equation2} as:

\begin{equation}
	v_\textrm{n} = \left( \frac{t_\textrm{p} - t_\textrm{l}}{l}  +  \frac{1}{c}  \right)^{-1} \,.
	\label{equation3}
\end{equation}

\section{Results and discussion}

\begin{figure}[h]
			\centering
			 \includegraphics[width=10cm]{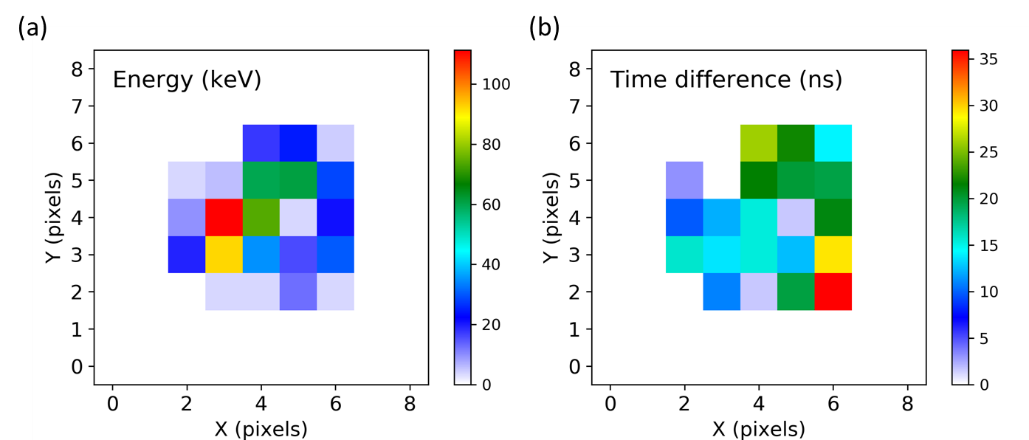}
			 \caption{Energy (a) and time (b) structure of a sample recoiled proton track registered by Timepix3 detector.} 
			\label{figure6}
\end{figure}

We used several criteria based on both the morphologic and deposited energy structure of registered tracks of the particles for the identification of recoiled protons' tracks. A sample recoiled proton track is shown in Figure~\ref{figure6}, where both the energy deposited in individual pixels and the time stamps values assigned to the pixels are shown.

\begin{figure}[h]
			\centering
			 \includegraphics[width=7cm]{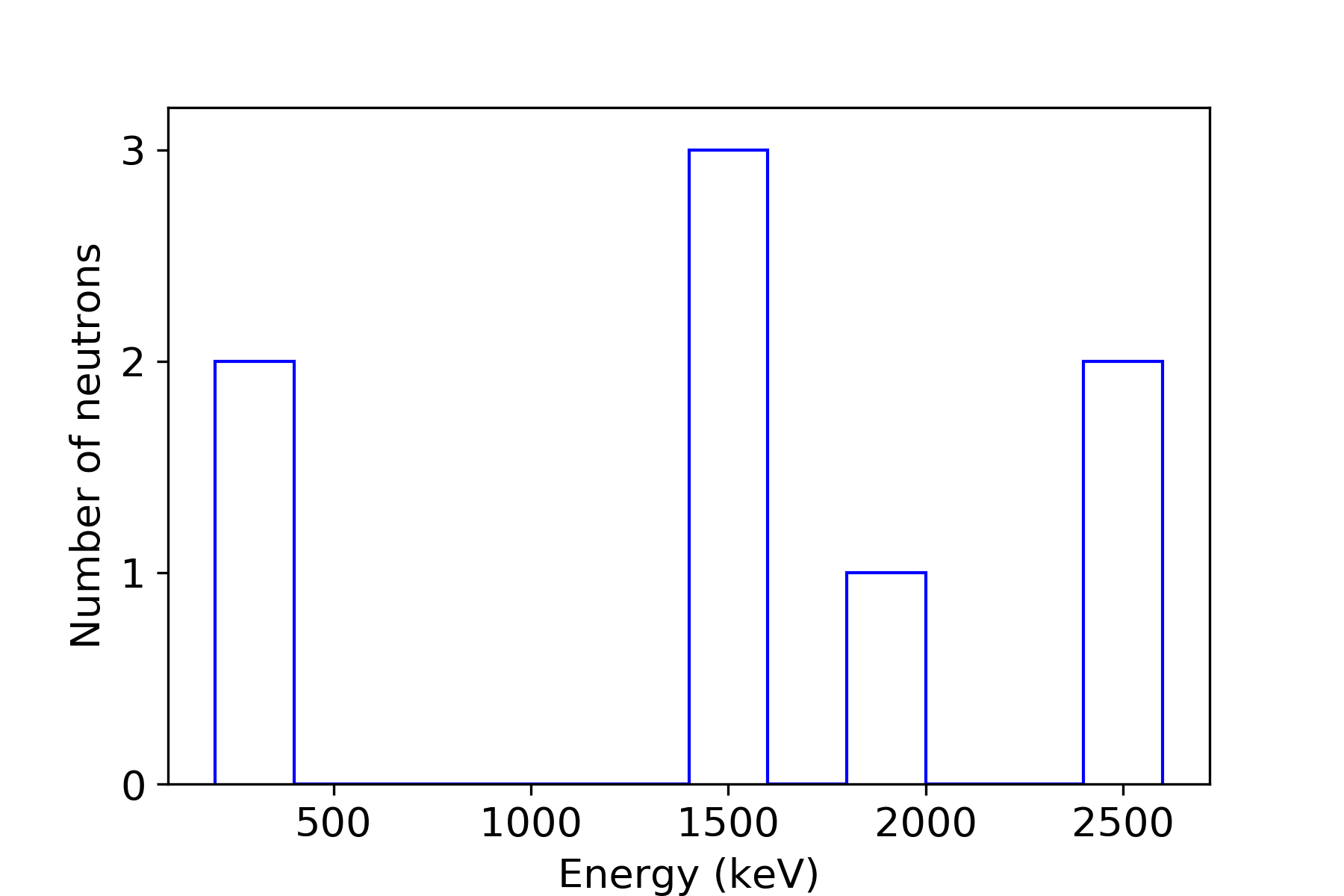}
			 \caption{Histogram of neutron energies determined through their measured TOF.} 
			\label{figure7}
\end{figure}

In total, we have identified 8 recoiled protons under the PE converter. For the correct determination of $t_\textrm{l}$, we took the first X-ray/electron track which was detected at the beginning of the bunch of the tracks during the laser shot. The TOF energy spectrum of the recoiled protons derived using equation~\ref{equation3} can be seen in Figure~\ref{figure7}, where width of energy bins and precision of TOF determination is taken into consideration.

From the spectrum seen in Figure~\ref{figure7} we can assume that we directly detected two D-D fusion neutrons with 2.45~MeV energy and six which underwent scattering either along the way to the detector or on the walls or the material inside of the experimental hall for the lowest apparent energies. Several arguments support this conclusion. First, we did not see any recoiled protons outside of the PE converter region. The threshold of neutron induced nuclear reactions in Si is $\sim4$~MeV.

\begin{figure}[h]
			\centering
			 \includegraphics[width=10cm]{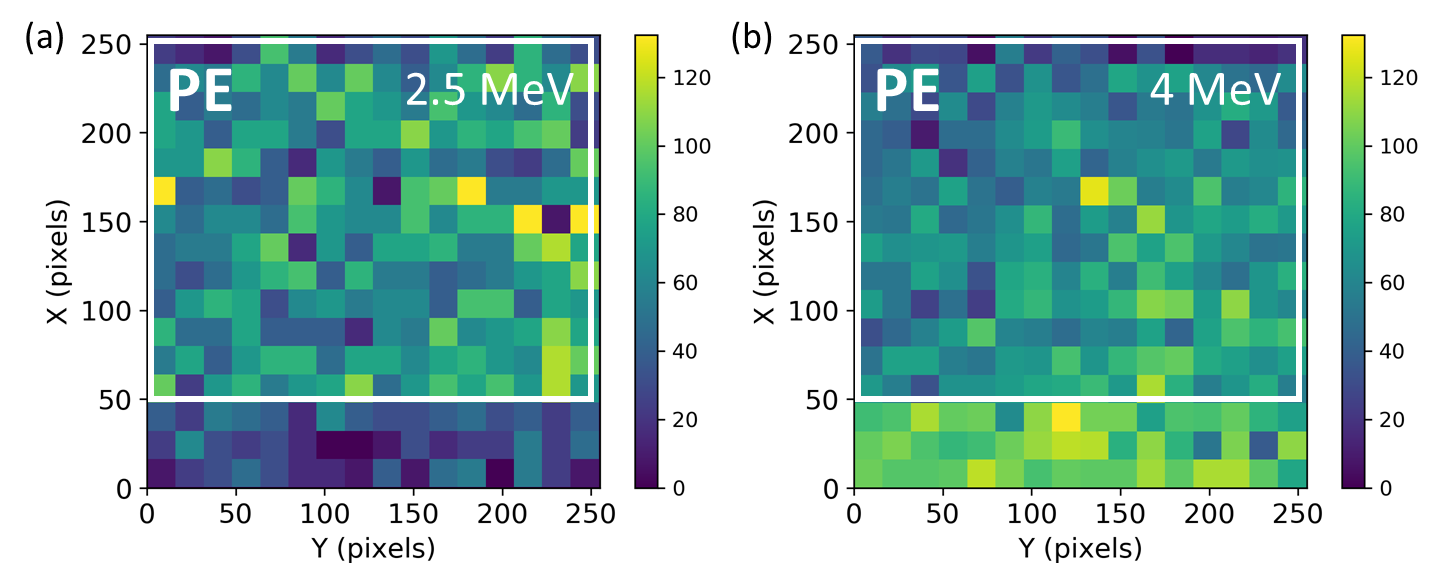}
			 \caption{Recoiled protons centroids map on Timepix3 detector covered with 1~mm thick PE converter. On panel~(a) results from measurement with 2.5~MeV neutrons generated with VdG accelerator can be seen, on panel~(b) results from experiment with 4~MeV neutrons.} 
			\label{figure8}
\end{figure}

\begin{figure}[h]
			\centering
			 \includegraphics[width=7cm]{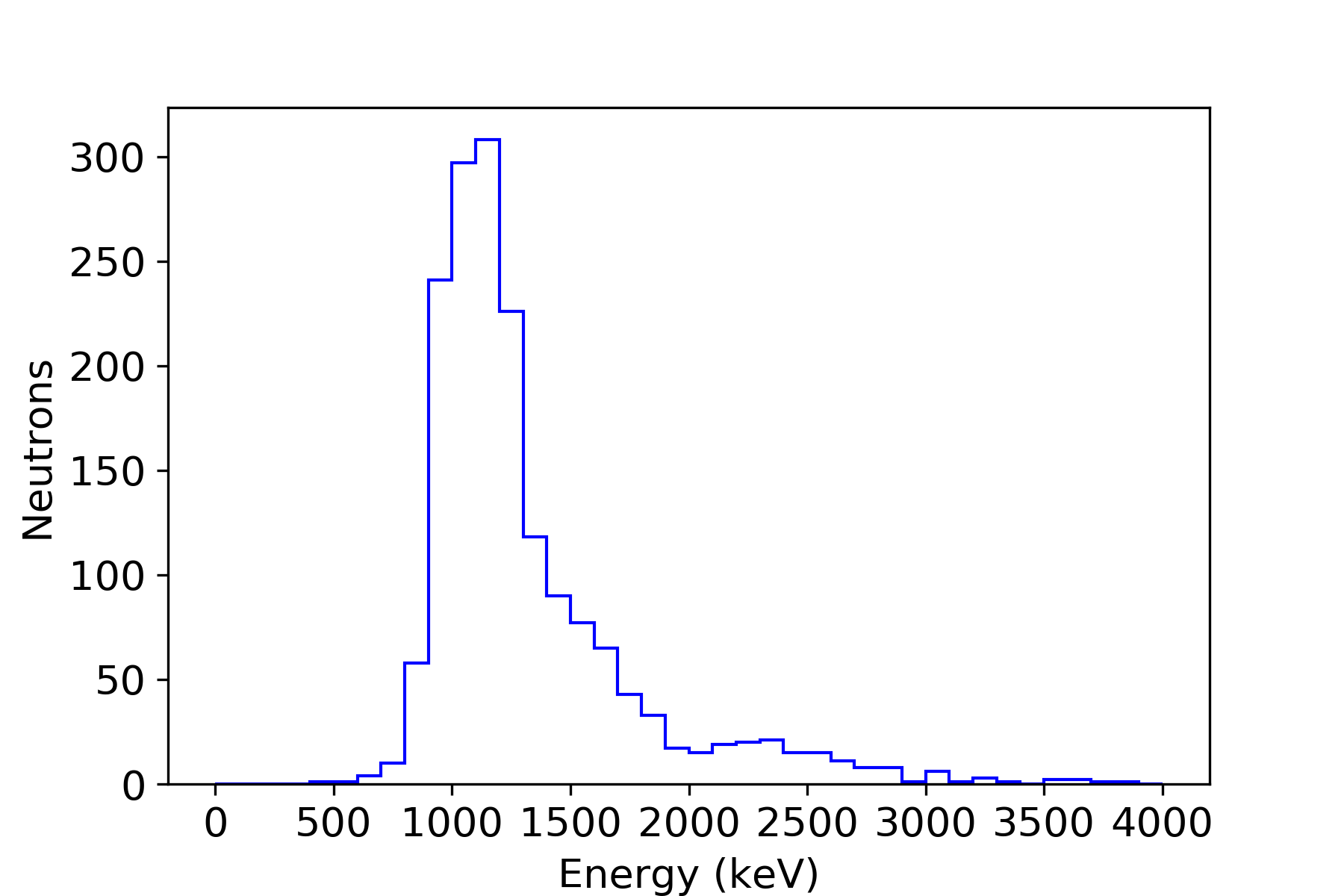}
			 \caption{The energy spectrum of recoiled protons from the measurement with 2.5~MeV neutrons generated with VdG accelerator.} 
			\label{figure9}
\end{figure}

\begin{figure}[h]
			\centering
			 \includegraphics[width=10cm]{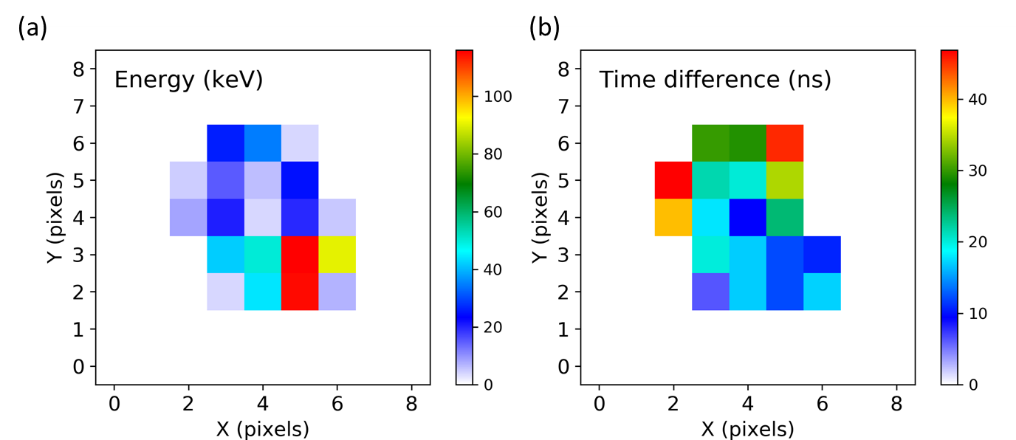}
			 \caption{Energy~(a) and time~(b) structure of a sample recoiled proton track from VdG accelerator registered by Timepix3 detector.} 
			\label{figure10}
\end{figure}

Secondly, we conducted an experiment with neutrons generated using Van de Graaff (VdG) type of accelerator at the Institute of Experimental and Applied Physics in Prague. We used the same Timepix3 detector as in the experiment at CLAPA facility, this time only covered with a 1~mm thick PE converter. We measured at two neutrons energies, 2.5~MeV and 4~MeV. A 2D map of the detector with marked PE converter region and recoiled protons centroids are shown in Figure~\ref{figure8} for both energies. The energy spectrum of the recoiled protons for 2.5~MeV neutrons is plotted in Figure~\ref{figure9}. We see that most of the recoiled protons have energy around 1.2~MeV, and that agrees with theory of elastic scattering. A sample recoiled proton track for 2.5~MeV neutrons is plotted in Figure~\ref{figure10}.

\begin{figure}[h]
			\centering
			 \includegraphics[width=7cm]{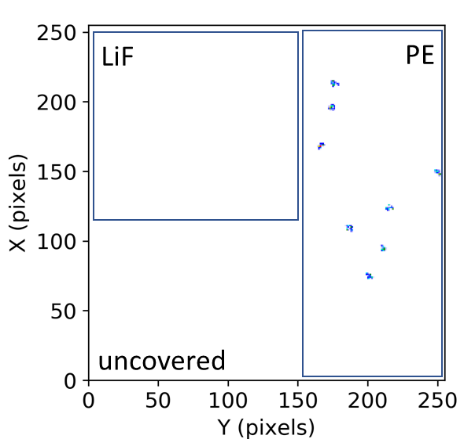}
			 \caption{The actual positions of the tracks of the identified recoiled protons. Positions of neutron converters are plotted as well.} 
			\label{figure11}
\end{figure}

One can see that neutrons with energy of 2.5~MeV (see panel~(a) of Figure~\ref{figure8}) are producing recoiled protons predominantly in the area covered by the PE converter. On the other hand, neutrons with 4~MeV energy are producing alphas and protons even in the uncovered area (see panel~(b) of Figure~\ref{figure8}). Similarities can be seen also when comparing recoiled proton track from the CLAPA experiment (figure~\ref{figure6}) with the VdG data (figure~\ref{figure10}). If we plot tracks identified as recoiled protons in the actual experiment (see Figure~\ref{figure11}), it can be clearly seen that all of them are in the area underneath the PE converter. This fact gives us good evidence that recoiled protons are created only by neutrons with energies lower than 4~MeV.

\begin{figure}[h]
			\centering
			 \includegraphics[width=7cm]{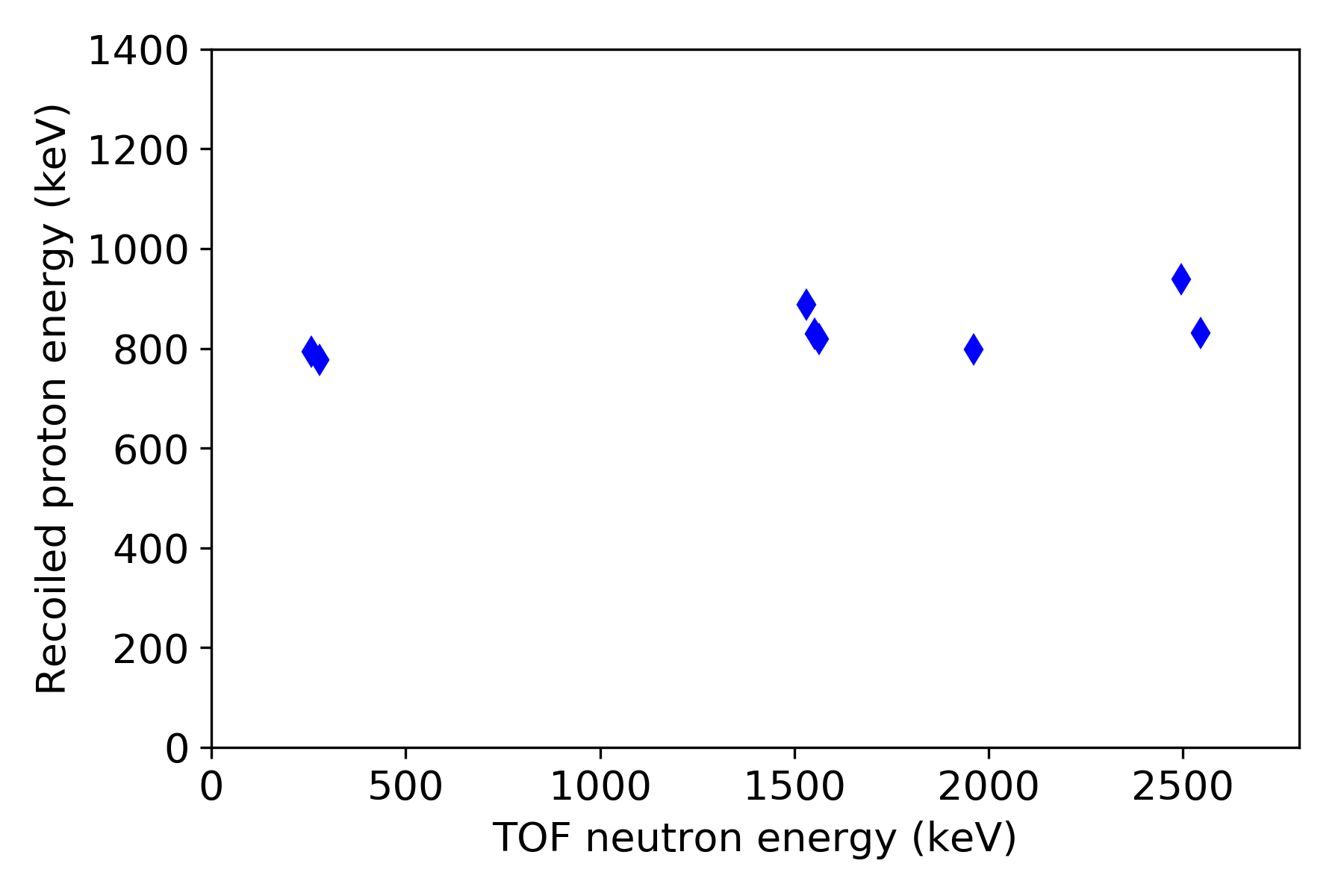}
			 \caption{Plot of neutron energies derived from TOF measurements versus recoiled protons energy measured directly by Timepix3 detector.} 
			\label{figure12}
\end{figure}

Another information can be deduced from the plot displayed in Figure~\ref{figure12}. The neutron energies derived from TOF measurements, even if conditions are not ideal (short base, multiple scattering, noise), are plotted against the measured recoiled protons energies. One can clearly see that two of the recorded neutrons induced events are 2.45~MeV fusion neutrons, while the recoiled proton energy is around 800~keV. This energy is lower than the energy of recoiled protons created by 2.5~MeV neutrons at VdG, which was  around 1.2~MeV. The difference is probably caused by a calibration offset during the CLAPA experiment. The rest of the neutrons lost part of their energy by scattering in the PE converter and in the material in the experimental hall. Two neutrons with low TOF derived energy (compared to the others) seen on the left of the plot are probably neutrons which traveled to opposite side of the laboratory and then back to the detector. In general, all the recoiled protons have about the same energy, thus we can assume that all of them belong to 2.45 MeV~neutrons. 

For the calculation of the efficiency of neutron generation we should only take the two neutrons with the highest TOF energy, since the neutrons did not scatter on the way to the detector. The distance between the detector and the target was 148~cm and 103~cm, respectively for those two neutrons. The detection efficiency for 2.45~MeV neutron detection is $3.6 \times 10^{-4}$ according to the calibration measurement in~\cite{Rubovic2018}. Assuming an isotropic angular distribution of the generated neutrons, we get $\left( 2.4 \pm 1.8 \right) \times 10^7$ generated neutrons per 1~J of laser energy. This finding is one order of magnitude higher than the one reported by Curtis et al.~\cite{Curtis2018}, who recently performed a similar experiment with CD$_2$ nano structured targets.

Using hybrid semiconductor pixel detectors brings several advantages over scintillation detectors, which are usually employed for such measurements. Firstly, it is possible to clearly discriminate neutron induced events from gamma background, as explained in this paper. Secondly, we can count each interacting neutron in this way, so the probability of neutron count underestimation is effectively minimized.

The precision and the efficiency of TOF measurement can be further improved in two ways. First, by using a thinner Si sensor, preferably with 100~$\mu$m thickness and by using a higher bias voltage. In this way, the uncertainties caused by drift time and absorption in different depths of the sensor can be partially overcome. Second, placing more detectors further (e.g.~four detectors at twice the distance) from the target would set a longer base while retaining the same geometrical efficiency. During the experiment we found that there is a risk of saturating the detector with laser light entering the sensor through the side of the sensor which is not metalized. This issue can be overcome by further improvements in the experimental setup, for example by light tight encapsulation of the detector assembly.. Such a device could possibly be used directly inside the vacuum experimental chamber in future experiments.

\section{Conclusion}	

We detected D-D fusion neutrons generated in a laser plasma using the hybrid semiconductor pixel detector Timepix3. The 2.45~MeV neutrons were detected via recoiled protons created in the PE converter placed directly on a 500~$\mu$m thick Si sensor. The energy of recoiled protons was measured directly by the Timepix3 detector and the energy of neutrons was determined from TOF measurements, where the starting point of the measurement was the detection of X-rays or electrons created by a laser shot. In total 8~neutrons were detected from 81~shots with the average on-target shot energy of 0.8~J. If we only take into account neutrons which did not scatter, it means that $\left( 2.4 \pm 1.8 \right) \times 10^7$ neutrons were generated for 1~J of incoming laser energy. Low number of detected neutrons is caused by very smal surface angle covered with the detector (1~cm$^2$ in the distance of 1~m). With these measurements, we prove that Timepix3 is a suitable tool for investigations under the difficult conditions (e.g.~high electro-magnetic noise) of laser experiments. 

\section*{Acknowledgement}

The support by following projects is acknowledged: European Regional Development Fund-Project ``Engineering applications of microworld physics'' (No. CZ.02.1.01/0.0/0.0/16\textunderscore019/0000766); National Natural Science Foundation of China (Grant No. 11775010); NSFC innovation group project (No. 11921006); National Grand Instrument Project (Grant No. 2019YFF01014402); Strategic Priority Research Program of the Chinese Academy of Sciences (Grant No. XDB16). The work was done in the frame of Medipix collaboration.



\end{document}